\documentclass[11pt]{article}
\usepackage{graphicx}
\usepackage{cite}

\newcommand{ \be}{\begin{equation}}
\newcommand{ \ee}{\end{equation}}
\newcommand{ \bea}{\begin{eqnarray}}
\newcommand{ \eea}{\end{eqnarray}}
\newcommand{ \mysmall}[1]{\scriptscriptstyle #1} 
\newcommand{ \smallmax}{{\rm\scriptscriptstyle max}} 
\newcommand{ \smallmin}{{\rm\scriptscriptstyle min}} 
\newcommand{ \mw}{M_{\mysmall{W}}}

\newcommand{ \gl}{g_{\mysmall{L}}}
\newcommand{ \gr}{g_{\mysmall{R}}}
\newcommand{ \fl}{f_{\mysmall{L}}}
\newcommand{ \fr}{f_{\mysmall{R}}}
\newcommand{ \flr}{f_{\mysmall{L} \mysmall{R}}}
\newcommand{ \fx}{f_{\mysmall{X}}}

\newcommand{ \nuov}{\bar{\nu}}

\newcommand{ \eq}[1]{eq.~(\ref{eq:#1})}

\newcommand{ \gev}  {\mbox{ GeV}}
\newcommand{ \mev}  {\mbox{ MeV}}
\newcommand{ \kev}  {\mbox{ keV}}

\newcommand\hepnumber{hep-ph/0412165}

\begin{document}

\begin{titlepage}
\begin{flushright}
  \hepnumber\\
\end{flushright}
\renewcommand{\thefootnote}{\fnsymbol{footnote}}

\begin{center}
\vspace{10mm} {\LARGE\bf Dynamical Zero in {\boldmath $\bar{\nu}_e$
\unboldmath}-- {\boldmath $e^-$\unboldmath} Scattering and \\[2mm] the
Neutrino Magnetic Moment}

\vspace{13mm}
{\large\bf      J.~Bernab\'eu$^a$, J.~Papavassiliou$^a$, and 
                M.~Passera$^{a,b}$}

\setcounter{footnote}{0}
\vspace{12mm}

{\it    $^a$ Departament de F\'\i sica Te\`orica and IFIC Centro Mixto,
        Universitat de Val\`encia--CSIC, E-46100, Burjassot, Val\`encia, 
	Spain \\[3mm] 

        $^b$ Dipartimento di Fisica ``G.~Galilei'', Universit\`{a} 
        di Padova and \\ 
	INFN, Sezione di Padova, I-35131, Padova, Italy}

\vspace{2cm}
{\large\bf Abstract} 
\end{center}

\vspace{3mm} 
\noindent 
The Standard Model differential cross section for $\bar{\nu}_e-e^-$ elastic
scattering vanishes exactly, at lowest order, for forward electrons and
incident $\bar{\nu}_e$ energy close to the rest energy of the electron.
This dynamical zero is not induced by a fundamental symmetry of the
Lagrangian but by a destructive interference between the left- and
right-handed chiral couplings of the electron in the charged and neutral
current amplitudes.  We show that lowest-order analyses based on this
favorable kinematic configuration are only mildly affected by the inclusion
of the $O(\alpha)$ radiative corrections in the $\bar{\nu}_e-e^-$
differential cross section, thus providing an excellent opportunity for the
search of ``new physics''.  In the light of these results, we discuss
possible methods to improve the upper limits on the neutrino magnetic moment
by selecting recoil electrons contained in a forward narrow cone. We
conclude that, in spite of the obvious loss in statistics, one may have a
better signal for small angular cones.

\end{titlepage}



\subsection*{Introduction}

One of the most important challenges of elementary particle physics today is
the detailed study of neutrino properties, such as neutrino masses and
mixings, the nature of massive neutrinos (Dirac or Majorana), and their
electromagnetic properties.  The possibility of a non-vanishing neutrino
magnetic moment has been the focal point of various investigations, because
its presence would provide a strong indication for physics beyond the
Standard Model ({\small{SM}}), given that within the {\small{SM}}, i.e.\
with massless neutrinos, it vanishes.  If the standard theory is extended to
include the right-handed neutrino field, and global lepton number symmetry
is enforced, the resulting Dirac neutrino with mass $m_{\nu}$ acquires a
magnetic moment \cite{Lee:1977ti} given by $\frac{3}{4 \sqrt{2} \pi^{2}}
G_{F} m_{\nu} m \mu_{B} \simeq 3.2 \times 10^{-19} (\frac{m_{\nu}}{\rm eV})
\mu_{B}$, where $\mu_{B} = e/2m$ is the electron Bohr magneton, and $m$ is
the electron mass. Given the current upper bound on the neutrino mass of
$m_{\nu} < 3$ eV~\cite{Eidelman:2004wy}, it follows that the ``standard''
contribution to the neutrino magnetic moment is less than $1\times\,10^{-18}
\mu_{B}$.  Such a small upper bound is far beyond the reach of any
present-day experiment.  However, there exist many models beyond the
standard theory in which the induced magnetic moment of neutrinos could be
several orders of magnitude larger (see, for example,~\cite{Babu:1989wn}).
The available bounds from terrestrial experiments and astrophysical
observations span a range of two orders of magnitude, $(10^{-10}-10^{-12})
\mu_{B}$ \cite{Eidelman:2004wy}.  It would clearly be very important to
establish new ways for providing more stringent experimental bounds for the
neutrino magnetic moment.

Motivated by this objective, in the present article we revisit the
``dynamical zeros'' appearing in the tree-level differential cross section
for antineutrino--electron elastic scattering ~\cite{SBBP93}. These zeros
are dynamical in the sense that they appear inside the physical region of
the kinematical variables describing the scattering and their location
depends on the fundamental parameters of the theory, as opposed to
kinematical zeros, which appear at the boundary of the physical region and
do not depend on dynamical parameters.  Obviously, any non-standard
contribution to the above cross section stemming from a neutrino magnetic
moment will have to compete against the standard one.  It has therefore been
proposed \cite{SBBP93} to exploit the vanishing of the SM cross section at
the aforementioned special kinematic configurations in order to expose the
possible effects associated with the neutrino magnetic moment. There are
three main issues which have forestalled the implementation of the
aforementioned strategy. First of all, given that these zeros are not
protected by any symmetry of the theory, there is no a priori reason why
they should survive higher-order corrections.  Second, their sensitivity to
the finite energy resolution needs to be established.  Third, it is not
clear at first sight whether what one gains in precision by selecting only
events displaying the zeros outways what one loses in statistics by
discarding all remaining events; this delicate balance could make the
practical usefulness of this method questionable.

In this paper we show that the inclusion of radiative corrections affects
the presence of the dynamical zeros only very mildly, and have therefore no
appreciable impact on the applicability of the dynamical zeros.  Moreover,
we demonstrate that the effect of the finite energy resolution leads to a
gradual smearing of the sharp ``dip'' which appears in the cross section in
the vicinity of the zero when perfect resolution is assumed.  Despite this
smearing, as shown in fig.~1, for realistic values of the energy resolution
one still obtains a clear suppression of the standard contribution compared
to that with a non-zero neutrino magnetic moment.  Finally, we argue that,
by selecting the recoil electrons contained in a forward cone centered
around the direction of the momentum of the incident neutrino, despite the
resulting loss in statistics, one can in fact improve on the existing bounds
on the neutrino magnetic moment.

\subsection*{Lowest Order}

Consider the elastic scattering $\nuov_l +e^- \rightarrow \nuov_l+e^-$
($l=e$, $\mu$, or $\tau$) in the frame of reference in which the electron is
initially at rest.  If we neglect terms of order $r/\mw^2$, where $r$
indicates any of the Mandelstam variables intervening in the scattering
process and $\mw$ is the $W$ boson mass, the lowest-order {\small SM}
prediction for this elastic differential cross section is, neglecting
antineutrino masses,~\cite{tH71}
\be
   \left[\frac{d\sigma}{dE}\right]_0 = \;
	\frac{2mG_{F}^2}{\pi} \left[\, \gr^2 +\gl^2 
   	\left(1-z\right)^2 -\gl \gr 
        \frac{m z}{E_{\nu}}\right].
\label{eq:LOdE}
\ee 
$G_{F}=1.16637(1) \times 10^{-5}\gev^{-2}$ is the Fermi coupling constant,
$m$ is the electron mass, $\gl = \sin^2 \!\theta_{\mysmall{W}} \pm 1/2$
(upper sign for $\nuov_e$, lower sign for $\nuov_{\mu,\tau}$), $\gr =
\sin^2\!\theta_{\mysmall{W}}$, and $\sin^2\!\theta_{\mysmall{W}} \approx
0.23$ is the squared sine of the weak mixing angle. In this elastic process
the electron recoil energy $E$ ranges from $m$ to $E_{\smallmax} =$ $[m^2
+(2E_{\nu} +m)^2]/[2(2E_{\nu} +m)]$, where $E_{\nu}$ is the incident
antineutrino energy. Also, $P=\sqrt{E^2-m^2}$ and $T=E-m$ are the final
electron three-momentum and kinetic energy, $z=T/E_{\nu}$, and $\cos\theta =
(1+m/E_{\nu})(T/P)$ is the cosine of the angle between the momenta of the
recoil electron and incident antineutrino. Note that eq.~(\ref{eq:LOdE})
is derived averaging over the polarizations of the initial-state electrons
and summing over their final-state helicities. The corresponding formula for
neutrino--electron scattering is simply obtained from \eq{LOdE} by
interchanging $\gl$ and $\gr$.

Some time ago the authors of ref.~\cite{SBBP93} showed the existence of
dynamical zeros in the helicity amplitudes for antineutrino--electron
elastic scattering at lowest order in the {\small SM}. These zeros are not
induced by a symmetry of the {\small SM} Lagrangian, but by a destructive
interference between left- and right-handed electron contributions to the
amplitudes.  Their location depends on the values of the fundamental
parameters $\gl$ and $\gr$. In particular, for scattering of electron
antineutrinos on electrons, the additional charged current contribution to
$\gl$ provides the appropriate cancellation with $\gr$.  On the other hand,
for scattering of electron neutrinos on electrons, the interference between
right and left couplings induced by charged and neutral current amplitudes
is constructive. The analysis of ref.~\cite{SBBP93} furnishes all the
information concerning dynamical zeros for both unpolarized and polarized
differential cross sections (the analytic formulae for the differential
cross sections of the elastic $\nu_l$--$e^-$ and $\nuov_l$--$e^-$
scatterings with all polarization states specified were computed
in~\cite{MP02}).  For reasons of experimental simplicity we will concentrate
here only on the unpolarized case, but refer the reader to~\cite{MP02,Polar}
for interesting opportunities offered by the study of polarization effects
in (anti)neutrino--electron scattering.

The differential cross section in \eq{LOdE} for $\nuov_{l}=\nuov_{e}$
vanishes exactly for antineutrino energy
\be 
     E_{\nu,0}= \frac{m}{2}\left( \frac{\gl}{\gr} -1\right) = 
     \frac{m}{4\sin^2\!\theta_{\mysmall{W}}}
\ee
and maximum corresponding electron recoil energy $E_{\smallmax}(E_{\nu,0})
\sim 5m/3$ (i.e., backward outgoing neutrino and forward electron).  We
should emphasize that $E_{\nu,0} \sim m$ lies inside the range of the
reactor antineutrino spectrum (in fact, it is around the peak -- see
fig.~2), and forward electrons with maximum recoil energy provide a
favorable kinematic configuration from the experimental point of view. Note
that there are no dynamical zeros in the unpolarized differential cross
section of $\nuov_{\mu,\tau}+e^- \rightarrow \nuov_{\mu,\tau} + e^-$, with
only neutral currents.  Following ref.~\cite{SBBP93}, the use of this
interesting kinematic configuration for the search for physics beyond the
{\small SM} has been advocated in a number of detailed
studies~\cite{SBBP94,Se97-MSSV99,BP03}.

Figure 1 shows various plots of the differential cross section for the
elastic $\nuov_e-e^-$ scattering at maximum electron recoil energy
$E_{\smallmax}(E_{\nu})$ as a function of the incident antineutrino energy
$E_{\nu}$. The dotted line labeled by ``0'' is the lowest-order prediction
provided by \eq{LOdE}. The dynamical zero for $E_{\nu,0}=
m/({4\sin^2\!\theta_{\mysmall{W}}}) \sim 0.55 \mev$ is clearly
recognizable. Consider now the average of the lowest-order differential
cross section in the endpoint region $E_{\smallmax}- \Delta E < E <
E_{\smallmax}$,
\be
      \overline{\left[\frac{d\sigma}{dE}\right]}_0 = \;
      \frac{1}{\Delta E} \int_{E_{\smallmax}- \Delta E}^{E_{\smallmax}} 
      \left[\frac{d\sigma}{dE}\right]_0 dE.
\label{eq:LOdEaverage}
\ee
This function of $E_\nu$ is plotted in fig.~1 for five different values of
the energy range $\Delta E =$ 1, 5, 10, 15 and 20 keV (dotted lines). If
$\Delta E$ is the energy interval corresponding to the experimental energy
resolution, these lines clearly show how the ``dip'' corresponding to the
dynamical zero gets increasingly filled up by the decrease in energy
resolution. The remaining lines in the figure will be discussed in the
following sections. Note that the atomic binding of the target electrons has
been assumed to be negligible because the dynamical zero occurs at
$E_{\nu,0} \sim 0.55 \mev$ and $T \sim 0.38\mev$, a very high value of the
electron energy when compared with its binding. However, we refer the reader
to ref.~\cite{FMS00} for a detailed study of this issue for targets
characterized by very high electron binding energies.

\subsection*{Radiative Corrections}

As we already pointed out earlier, the dynamical zero of the elastic
$\nuov_e +e^- \rightarrow \nuov_e+e^-$ scattering is not protected by any
symmetry of the {\small SM} Lagrangian and radiative corrections could
significantly modify the lowest-order analysis presented in the previous
section. Moreover, the zero occurs at the endpoint of the electron spectrum
-- an exceptional kinematic configuration, as we will now discuss.  To
address $O(\alpha)$ corrections, a few considerations are in order. As we
mentioned earlier, we neglect terms of order $r/\mw^2$. Within this
approximation, which is excellent for present experiments, the $O(\alpha)$
corrections to this process can be naturally divided into two classes. The
first, which we will call ``{\small QED}'' corrections, consists of the
photonic radiative corrections that would occur if the theory were a local
four--fermion Fermi theory rather than a gauge theory mediated by vector
bosons; the second, which we will refer to as the ``electroweak'' ({\small
EW}) corrections, will be the remainder. The split-up of the {\small QED}
corrections is sensible as they form a finite (both infrared and
ultraviolet) and gauge-independent subset of diagrams. We refer the reader
to ref.~\cite{Si78-80} for a detailed study of this separation. The {\small
QED} corrections were first studied in the 1960s in the pioneering articles
of Lee and Sirlin~\cite{LS64}, and Ram~\cite{Ram67} in the framework of an
effective four-fermion V--A theory, and further investigated in several
subsequent articles~\cite{QED, Pa00}. We will use the complete results for
the {\small QED} corrections to the final electron spectrum which became
available only a few years ago~\cite{Pa00}.  The {\small EW} corrections
were computed by many authors~\cite{EW,BKS95}; we will employ the compact
expressions of ref.~\cite{BKS95}.

The {\small SM} prediction for the differential cross section $ \nuov_l + e
\rightarrow \nuov_l + e \;(+\gamma)$, where $(+\gamma)$ indicates the
possible emission of a photon, can be cast, up to corrections of
$O(\alpha)$, in the following form:
\bea
   \left[\frac{d\sigma}{dE}\right]_{\rm SM}
   & \!\!\!\!\!\!\!\! = & \!\!\!\!\!\! \frac{2mG_{F}^2}{\pi} 
        \Biggl\{\gr^2(E) \left[1+\frac{\alpha}{\pi} \fl(E,E_{\nu}) \right]
        +\gl^2(E) \left(1-z\right)^2 
        \left[1+\frac{\alpha}{\pi} \fr(E,E_{\nu}) \right] \nonumber\\
   & &  -\gl(E) \gr(E) \left(\frac{m z}{E_{\nu}}\right)
        \left[1+\frac{\alpha}{\pi} \flr(E,E_{\nu}) \right] \Biggr\}.
\label{eq:SMdE}
\eea 
The functions $\fx(E,E_{\nu})$ ({\small $X=L,R$ or $LR$}) describe the
{\small QED} effects of real and virtual photons~\cite{Pa00}, while the
deviations of the functions $\gl(E)$ and $\gr(E)$ from the lowest-order
values $\gl$ and $\gr$ reflect the effect of the electroweak
corrections~\cite{BKS95}.

As it was noted in refs.~\cite{Ram67,BKS95}, the $\fx(E,E_{\nu})$ functions
contain a term which diverges logarithmically at the end of the spectrum,
i.e.\ for $E=E_{\smallmax}$, which is precisely the kinematic configuration
required for the vanishing at $E_{\nu,0}$ of the lowest-order differential
cross section in \eq{LOdE}.  This feature, related to the infrared
divergence, is similar to the one encountered in the {\small QED}
corrections to the $\mu$--decay spectrum~\cite{BFS56,KS59}. If $E$ gets very
close to the endpoint we have $(\alpha/\pi)\fx(E) \sim -1$, clearly
indicating a breakdown of the perturbative expansion and the need to
consider multiple-photon emission. However, this divergence can be easily
removed, in agreement with the {\small KLN} theorem~\cite{KS59,KLN}, by
integrating the differential cross section over small energy intervals
corresponding to the experimental energy resolution, as we did in
\eq{LOdEaverage} for the lowest-order prediction,
\be
      \overline{\left[\frac{d\sigma}{dE}\right]}_{\rm SM} = \;
      \frac{1}{\Delta E} \int_{E_{\smallmax}- \Delta E}^{E_{\smallmax}} 
      \left[\frac{d\sigma}{dE}\right]_{\rm SM} dE.
\label{eq:SMdEaverage}
\ee

We are thus ready to assess the impact of the $O(\alpha)$ corrections on the
dynamical zero of the lowest-order differential cross section.  The solid
lines in fig.~1 represent, as a function of $E_{\nu}$, the average of the
{\small SM} differential cross section in the endpoint region
$E_{\smallmax}- \Delta E < E < E_{\smallmax}$ up to corrections of
$O(\alpha)$ (i.e., \eq{SMdEaverage}).  As for the lowest-order dotted lines
(see previous section), the label next to each solid line indicates one of
the five values of the energy range $\Delta E =$ 1, 5, 10, 15 and 20 keV.
There is no solid line labeled ``0'', as $[d\sigma/dE]_{\rm {\mysmall SM}}$
is not defined at the endpoint $E=E_{\smallmax}$.  Comparing each dotted
line for the lowest-order prediction with the corresponding solid one
inclusive of $O(\alpha)$ corrections, we conclude that, in all cases
considered, the effect of the lowest-order dynamical zero is only mildly
influenced by the inclusion of radiative corrections\footnote{ 
This analysis is based on the $O(\alpha)$ electron spectrum of \eq{SMdE},
which includes the bremsstrahlung radiation (real photons) emitted in the
scattering process. In bremsstrahlung events, however, some detectors do not
measure the electron energy $E$ separately, but only a combination of $E$
and the energy of the photon (see~\cite{Pa00} for a detailed study of this
issue). For this reason, we repeated our analysis using the QED corrections
of ref.~\cite{Pa00} appropriate for detectors measuring the total combined
energy of the recoil electron and the possible accompanying photon. Although
different from those appearing in \eq{SMdE}, also these corrections have no
appreciable influence on the effect of the lowest-order dynamical zero.}.
This relative stability under radiative corrections of the effect of the
dynamical zero provides solid foundations to all previous analyses based on
this favorable kinematic configuration.

\subsection*{Neutrino Magnetic Moment}

The dynamical zero of the {\small SM} differential cross section in
\eq{LOdE} provides an excellent opportunity to unveil or constrain ``new
physics'' effects. In particular, refs.~\cite{SBBP93,Se97-MSSV99} advocated
the possibility of employing it to search for a neutrino magnetic moment.

If neutrino masses are neglected, a neutrino magnetic moment of magnitude
$\mu_{\nu}\mu_{B}$ increases the {\small SM} differential cross section for
the elastic scattering $\nuov_e +e^- \rightarrow \nuov_e+e^-$ by~\cite{K84}
\be
   \left[\frac{d\sigma}{dE}\right]_{\rm M}= \;
	\frac{\pi \alpha^2 \mu_{\nu}^2}{m^2} \left[\, 
	  \frac{1}{T} - \frac{1}{E_{\nu}} \right].
\label{eq:MdE}
\ee   
The measurement of a recoil differential spectrum larger than expected could
thus signal the existence of a neutrino magnetic moment, especially if it is
characterized by the distinctive low-energy $1/T$ enhancement. To this end,
it is important to minimize the detection threshold for the electron recoil
energy -- a difficult task, given the generally increasing detector
background with decreasing energy. On the other hand, as first pointed out
in ref.~\cite{SBBP93}, rather than looking for regions of lowest possible
energies where the differential cross section in \eq{MdE} becomes large
enough to be comparable with the {\small SM} spectrum of \eq{LOdE}, one can
take advantage of the dynamical zero of the latter. This can be immediately
appreciated by looking at the three dashed lines in fig.~1, which represent
$[d\sigma/dE]_{\rm {\mysmall M}}$ at maximum electron recoil energy
$E_{\smallmax}(E_{\nu})$ for three different values of the neutrino magnetic
moment: $\mu_{\nu}=1.0 \times 10^{-10}$, which is the current experimental
90\% {\small CL} upper bound by the {\small MUNU}
collaboration~\cite{Munu03}, $0.5 \times 10^{-10}$, and $0.2 \times
10^{-10}$. Electron antineutrinos with energy around $E_{\nu,0}$ could
therefore provide the possibility to study low values of $\mu_{\nu}$.  This
interesting conclusion was reached in ref.~\cite{SBBP93} studying the
lowest-order {\small SM} cross section and, as we showed in the previous
section, the analysis of that reference is only mildly modified by the
inclusion of radiative corrections. For this reason, in the remaining part
of this article we will simplify our analysis by employing the lowest-order
{\small SM} cross section given by \eq{LOdE}, instead of the one inclusive
of $O(\alpha)$ corrections, \eq{SMdE}.

Antineutrinos of incident energy $E_{\nu,0}$ can be selected from a
continuous spectrum source by measuring both energy and direction of the
electrons recoiling from the elastic scattering. This can be realized with
detectors like the one of {\small MUNU}, an experiment carried out at the
Bugey nuclear power reactor, designed to study $\nuov_e-e^-$ elastic
scattering at low energy~\cite{Munu03, Munu}.  The differential cross
section measured at reactor antineutrino experiments can be compared with
the theoretical prediction given in terms of
\be
      \left\langle\frac{d\sigma}{dE}\right\rangle_{\rm TH} = \;
      \int_{E_{\nu,\smallmin} (E)}^\infty \lambda(E_{\nu})
      \left[\frac{d\sigma}{dE}\right]_{\rm TH} dE_{\nu},
\label{eq:THdEconv1}
\ee
where the subscript {\small ``TH''} stands for {\small ``0''} or {\small
``M''} and $\lambda(E_{\nu})$ is the normalized antineutrino spectrum
incident at the detector. $E_{\nu,\smallmin} (E)=(T+P)/2$ is the minimum
$E_{\nu}$ required to produce an electron with recoil energy $E$. If the
recoil angle $\theta$ can also be measured, then the analysis can be
restricted to events with electrons recoiling in the forward cone $\cos
\theta \geq \cos (\theta_\smallmax) \equiv \delta$ (note that for a given
value of $E$, the recoil electrons are restricted by kinematics to lie in
the cone $T/P<\cos \theta \leq 1$). Instead of \eq{THdEconv1}, the
theoretical prediction to match this selective measurement is
\be 
      \left\langle\frac{d\sigma(\delta)}{dE}\right\rangle_{\rm TH} =
      \; \int_{E_{\nu,\smallmin} (E)}^{E_{\nu,\smallmax} (E,\delta)}
      \lambda(E_{\nu}) \left[\frac{d\sigma}{dE}\right]_{\rm TH} dE_{\nu},
\label{eq:THdEconv2}
\ee
where the upper limit of integration is now $E_{\nu,\smallmax}
(E,\delta)=mT/(P\delta-T)$ and $T/P \leq \delta \leq 1$. Clearly, $\langle
d\sigma/dE\rangle_{\rm {\mysmall TH}}$ is the limit of $\langle
d\sigma(\delta)/dE\rangle_{\rm {\mysmall TH}}$ for $\delta \rightarrow
(T/P)^+$.  For a given value of $E$, we can thus select incident
antineutrinos with energies $E_{\nu,\smallmin} (E) \leq E_{\nu} \leq
E_{\nu,\smallmax} (E,\delta)$ by rejecting events with $T/P<\cos \theta <
\delta$. In particular, incident antineutrinos with $E_{\nu} \sim E_{\nu,0}$
can be selected by considering only $E \sim 5m/3$ electrons in a small
forward cone $\cos \theta \sim 1$.

To study the sensitivity of the theoretical prediction to the neutrino
magnetic moment when the recoil electrons are contained in the forward cone
$\cos\theta\geq \delta$, we introduce the ratio
\be
      S(E,\mu_{\nu},\delta)=
      \frac{\left\langle\frac{d\sigma(\delta)}{dE}\right\rangle_{\rm 
	  \mysmall{M}}}
	   {\left\langle\frac{d\sigma(\delta)}{dE}\right\rangle_{\rm 0}}. 
\label{eq:S}
\ee
This ratio is plotted in fig.~3 as a function of $T$ for three different
values of $\mu_{\nu}$. Solid lines indicate the ratio $S$ for
$\theta_\smallmax=10^{\circ}$, while dotted lines, labeled by ``no cuts'',
represent the same ratio obtained without restricting the angle of the
recoiling electrons (i.e., $\delta \rightarrow (T/P)^+$). The normalized
reactor antineutrino spectrum of ref.~\cite{KMS97} has been used for
$\lambda(E_{\nu})$, see fig.~2. For $\mu_{\nu}=1.0 \times 10^{-10}$, fig.~3
shows that the value of $T$ for which the convoluted magnetic moment cross
section becomes equal to the {\small SM} one $(S=1)$ moves from $\sim 300
\kev$ to $\sim 1200 \kev$ if the recoil electrons are contained in a
$10^{\circ}$ forward cone. The same angular restriction shifts from $\sim
150 \kev$ to $\sim 900 \kev$ the value of $T$ for which the magnetic moment
cross section for $\mu_{\nu}=0.5 \times 10^{-10}$ equals 50\% of the {\small
SM} one. Also from fig.~3 it can be seen that while at 250 keV $\langle
d\sigma/dE\rangle_{\rm \mysmall{M}}$ for $\mu_{\nu}=0.2 \times 10^{-10}$ is
only about 5\% of $\langle d\sigma/dE\rangle_0$, by discarding events with
$\theta>10^{\circ}$ the two cross sections become almost equally large.  The
upshot is that {\em angular cuts on the final-state electrons move the
$\mu_{\nu}$ sensitivity to higher (and thus more accessible) electron
energies.} We should point out that once the electron energy is fixed, the
larger the value of $\delta$, the smaller the upper bound
$E_{\nu,\smallmax}$ in the definition of \eq{THdEconv2}. This implies that
increasing the value of $\delta$ also increases the relative systematic
uncertainty of the convoluted cross section due to the lack of precise
knowledge of the low energy part of the reactor antineutrino spectrum. On
the other hand, the great advantage of going to regions where $S$ is large
is that the sensitivity to this systematic uncertainty is dramatically
diminished~\cite{LW01}.

This remarkable opportunity to overcome serious experimental limitations in
the search for neutrino magnetic moments, such as high backgrounds at low
energies, must certainly be confronted with the loss in statistics induced
by the angular selection. As the sensitivity to $\mu_{\nu}$ in direct search
experiments scales as $1/ \sqrt{N}$, where $N$ is the number of signal
events, we introduce the function $\bar{S}$ defined by
\be \bar{S}(E,\mu_{\nu},\delta)= S(E,\mu_{\nu},\delta)
      \left[\,\left\langle\frac{d\sigma(\delta)}{dE}\right\rangle_{\rm M}
      \!\! + \,
      \left\langle\frac{d\sigma(\delta)}{dE}\right\rangle_0\,\right]^{1/2}.
\label{eq:Sbar}
\ee
The square root expression multiplying $S$ takes into account the loss of
statistical accuracy caused by the rejection of events with $\cos \theta
<\delta$.  A figure of merit is provided in fig.~4, where the ratio $R =
\bar{S}(E,\mu_{\nu},\cos(\theta_\smallmax))/
\bar{S}(E,\mu_{\nu},\cos(45^{\circ}))$ (as an example, the value
$\theta_\smallmax=45^{\circ}$ employed by the {\small MUNU} Collaboration
has been chosen as reference) is plotted for $\theta_\smallmax=5^{\circ}$
and $10^{\circ}$. Figure 4 indicates that high values of the sensitivity
$S$, due to the selection of electrons contained in a narrow forward cone,
can overcompensate the loss in statistics induced by this angular selection,
provided the cone can be narrowed sufficiently.  Therefore, apart from the
previously discussed opportunities to reduce systematic uncertainties, the
search for $\mu_{\nu}$ can actually benefit from an angular restriction even
from the purely statistical analysis illustrated in fig.~4. Indeed, fig.~4
shows that for $\theta_\smallmax=10^{\circ}$ and $\mu_{\nu}=1.0 \times
10^{-10}$ (the current upper bound), $R$ is larger than one for $T$ in the
range $\sim 0.15-0.50\mev$. Higher ratios are obtained for
$\theta_\smallmax=5^{\circ}$, in a slightly higher energy range. Following
the conclusions of our previous paragraph on the benefits of the angular
selection in overcoming systematic limitations, we will finally note that
even analyses with $R<1$, although statistically disfavoured, might still
provide a better opportunity to search for $\mu_{\nu}$ than employing a
large $\theta_\smallmax = 45^{\circ}$ cone. Dedicated experimental
studies, with realistic backgrounds and systematic uncertainties, should 
address this delicate issue and make an analysis with real data.

\subsection*{Conclusions}

It is known that, due to a destructive interference between charged and
neutral current amplitudes, a dynamical zero appears in the lowest-order
{\small SM} cross section describing the $\bar{\nu}_e-e^-$ elastic
scattering.  In this article we studied several main issues related to the
applicability of this dynamical zero as a method for improving the bounds on
the values of the neutrino magnetic moment. In particular, by means of a
detailed analysis we demonstrated that the lowest-order dynamical zero
remains essentially unaffected by the inclusion of the $O(\alpha)$ radiative
corrections and, for realistic values of $\Delta E$, finite energy
resolution still allows for the isolation of possible ``new physics''
contributions related to the presence of a non-standard neutrino magnetic
moment.

Having established the persistence of the dynamical zero effects under
radiative corrections, we proceeded to argue that the experimental isolation
of events near the region of this special configuration is in fact
advantageous, despite the obvious loss in statistics.  The overcompensating
factor originates from the fact that, when the corresponding angular cuts
are imposed on the final-state electrons, the sensitivity to $\mu_{\nu}$
increases and moves to higher, and therefore more accessible, electron
energies.  A high sensitivity to $\mu_{\nu}$ also diminishes the sensitivity
to systematic uncertainties. In addition, for discussed values of
$\mu_{\nu}$, the signal function $\bar{S}$, defined in \eq{Sbar} to take
into account the loss in statistics, is larger, in a specific range of the
recoil electron energy, for a small $\theta_{\smallmax}\sim
5^{\circ}-10^{\circ}$ angular cone rather than for a large
$\theta_{\smallmax}\sim 45^{\circ}$ one.  Our results suggest that an
analysis with real data is worth to be done.

\subsection*{Acknowledgments}

It is a pleasure to thank C.~Broggini and P.~Minkowski for very useful
discussions. This work was supported by the MECyD fellowship SB2002-0105,
the MCyT grant FPA2002-00612, and by the European Union under contract
HPRN-CT-2000-00149.

\begin{figure}[tbp]
\vspace{-3cm}\hspace{0cm}\includegraphics[width=14cm,angle=0]{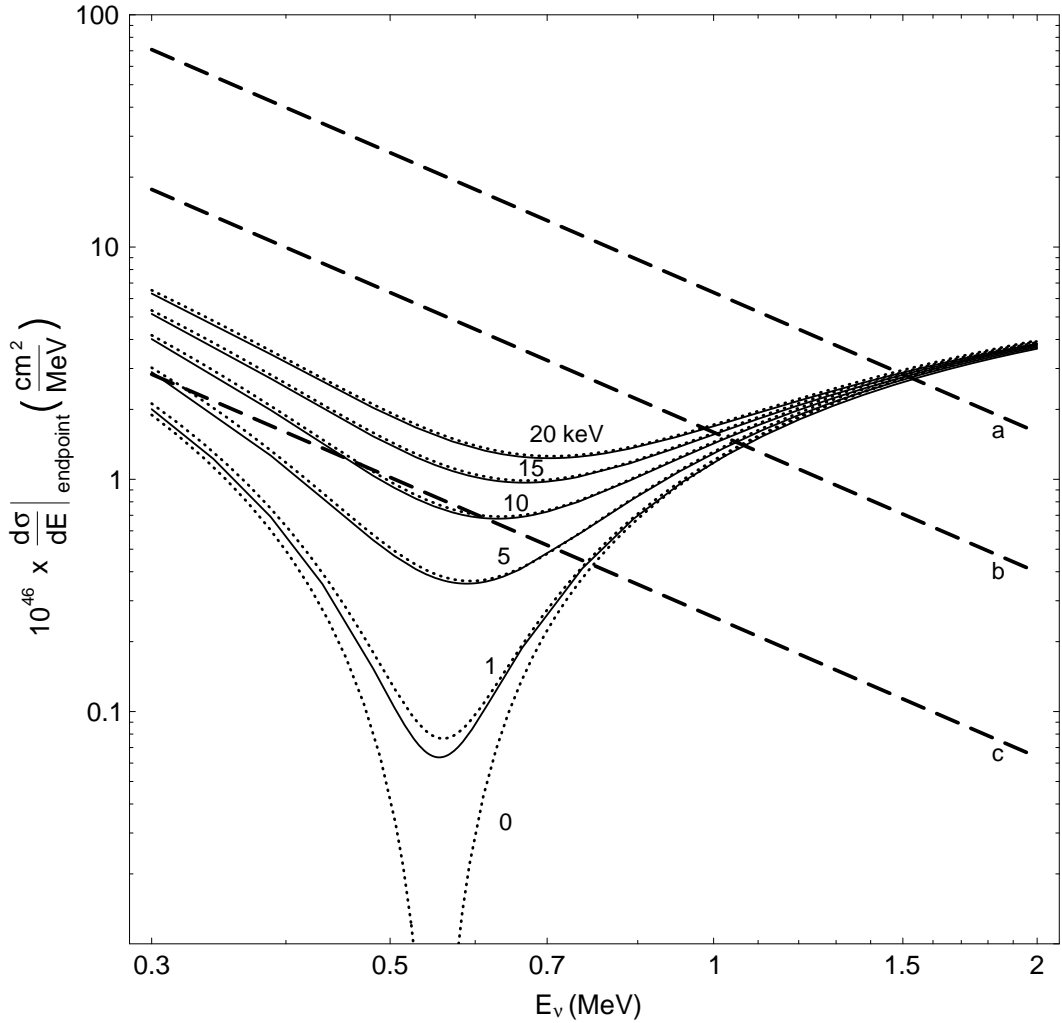}
\vspace{-2cm}
\caption{Differential cross section for the elastic $\nuov_e-e^-$ scattering
at maximum electron recoil energy $E_{\smallmax}(E_{\nu})$ as a function of
the incident antineutrino energy $E_{\nu}$. The dotted curve labeled by
``0'' indicates $[d\sigma/dE]_0$, the lowest-order {\small SM} prediction of
\eq{LOdE}. The dotted (solid) curves labeled by ``1''--``20 keV'' represent
the average $\overline{[d\sigma/dE]}_0$ ($\overline{[d\sigma/dE]}_{\rm
{\mysmall SM}}$) in the endpoint region $E_{\smallmax}- \Delta E < E <
E_{\smallmax}$ according to \eq{LOdEaverage} (\eq{SMdEaverage}). The labels
provide the value of $\Delta E$. The magnetic moment contribution
$[d\sigma/dE]_{\rm {\mysmall M}}$ is depicted by the three dashed lines
labeled ``a'', ``b'' and ``c'' for $\mu_{\nu}=1.0 \times 10^{-10}$, $0.5
\times 10^{-10}$ and $0.2 \times 10^{-10}$, respectively.}
\label{figure:figure1}
\end{figure}

\begin{figure}[tbp]
\vspace{-3cm}\hspace{0cm}\includegraphics[width=14cm,angle=0]{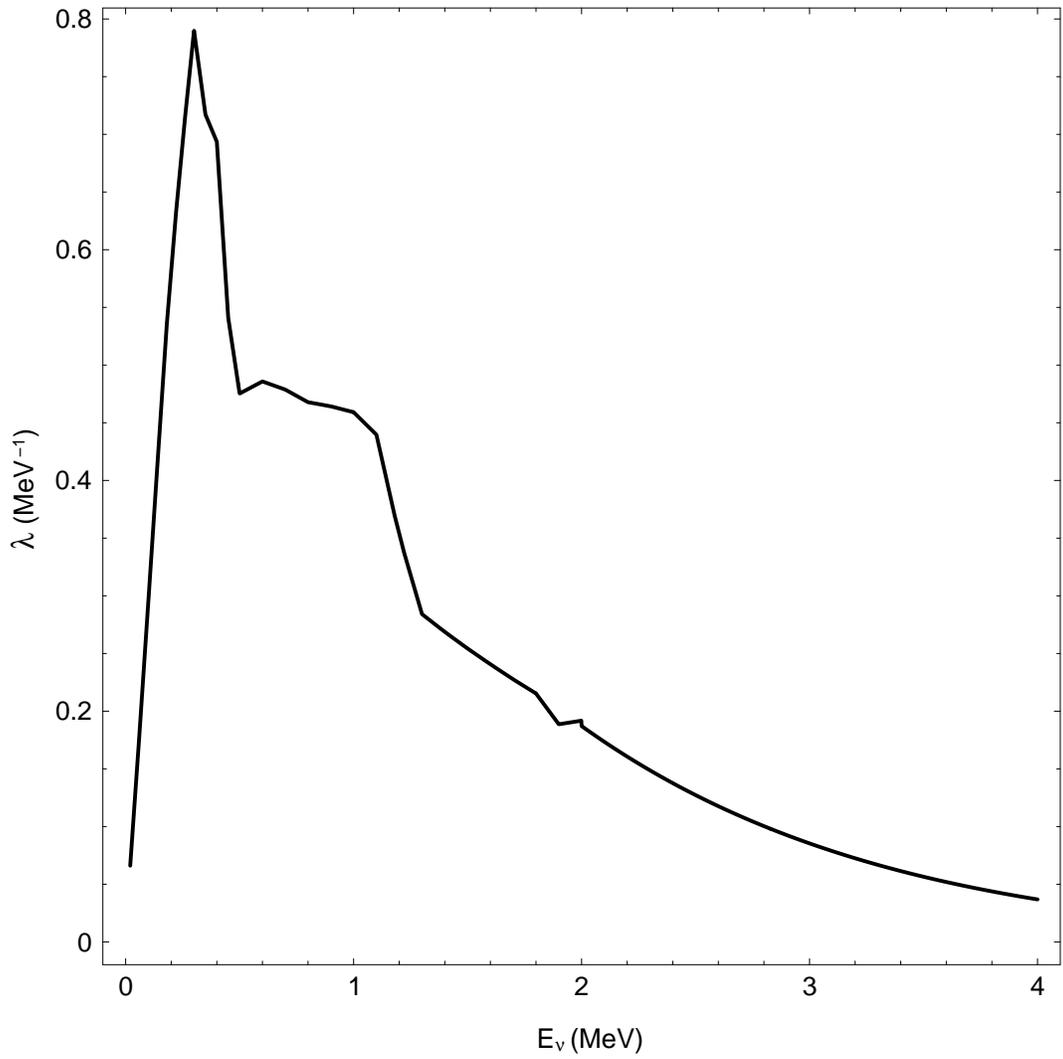}
\vspace{-2cm}
\caption{The normalized reactor antineutrino spectrum of
  ref.~\cite{KMS97}. Points for $E_{\nu} \leq 2$ MeV were interpolated
  linearly.}
\label{figure:figure2}
\end{figure}

\begin{figure}[tbp]
\vspace{-3cm}\hspace{0cm}\includegraphics[width=14cm,angle=0]{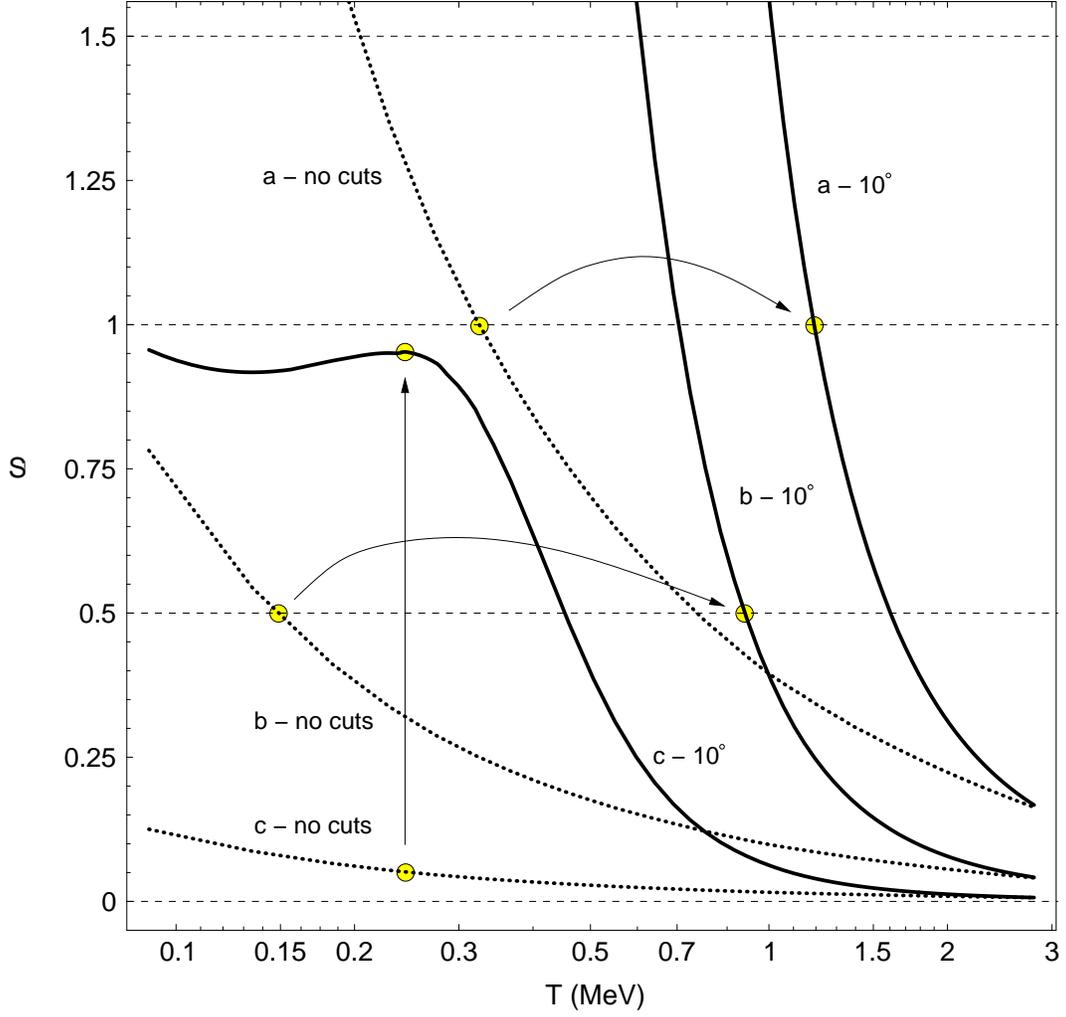}
\vspace{-2cm}
\caption{The ratio $S(E,\mu_{\nu},\delta)=\langle
d\sigma(\delta)/dE\rangle_{\rm {\mysmall M}}/\langle d\sigma(\delta)/dE
\rangle_0$ (\eq{S}) as a function of $T=E-m$. Solid lines are plotted for
$\theta_\smallmax=10^{\circ}$, while dotted lines, labeled by ``no cuts'',
represent the same ratio obtained without restricting the angle of the
recoiling electrons (i.e., $\delta \rightarrow T/P$).  As in fig.~1, the
labels ``a'', ``b'' and ``c'' stand for $\mu_{\nu}=1.0 \times 10^{-10}$,
$0.5 \times 10^{-10}$, and $0.2 \times 10^{-10}$, respectively.}
\label{figure:figure3}
\end{figure}

\begin{figure}[tbp]
\vspace{-3cm}\hspace{0cm}\includegraphics[width=14cm,angle=0]{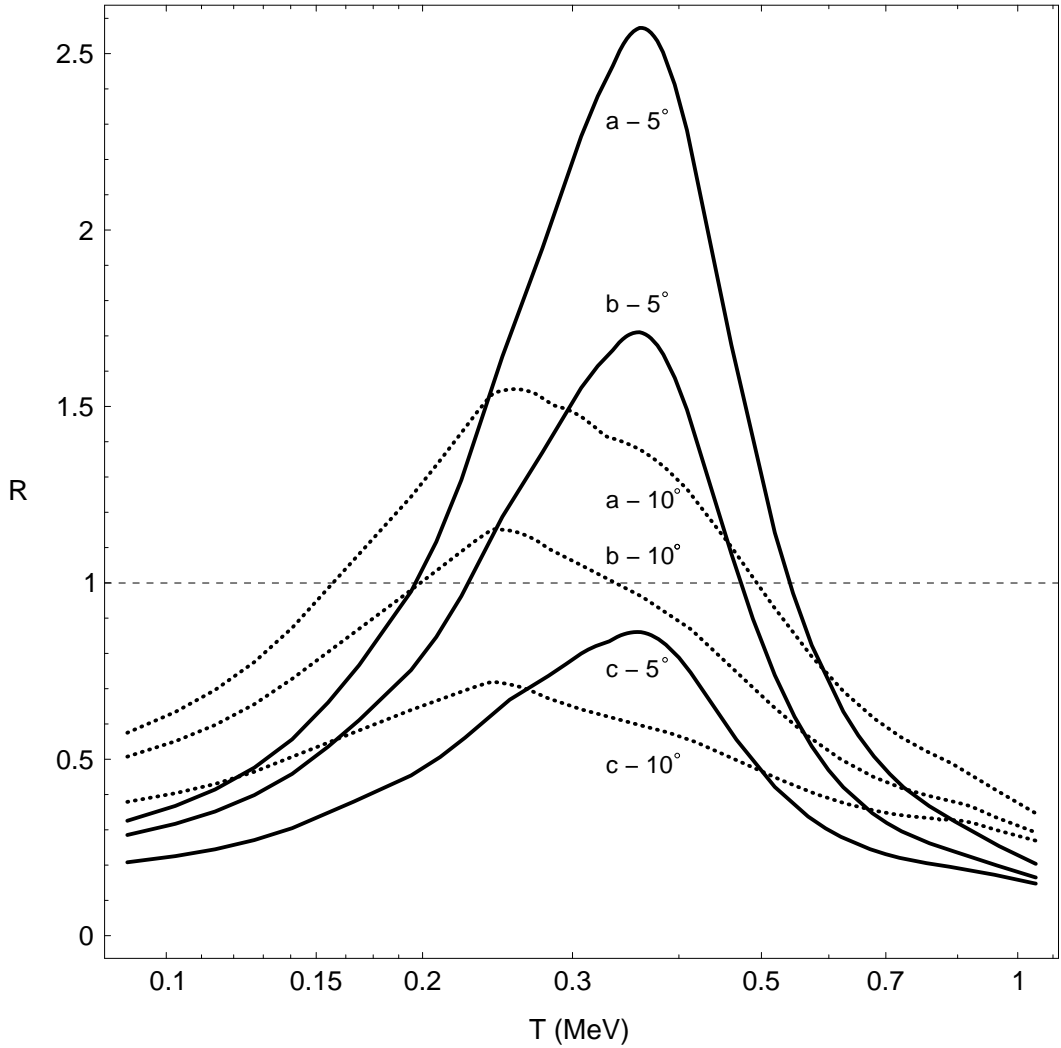}
\vspace{-2cm}
\caption{The ratio $R = \bar{S}(E,\mu_{\nu},\cos(\theta_\smallmax))/
\bar{S}(E,\mu_{\nu},\cos(45^{\circ}))$ as a function of $T=E-m$. Solid
(dotted) lines are plotted for $\theta_\smallmax=5^{\circ} ~(10^{\circ})$.
As in figs.~1 and 3, the labels ``a'', ``b'' and ``c'' stand for
$\mu_{\nu}=1.0 \times 10^{-10}$, $0.5 \times 10^{-10}$, and $0.2 \times
10^{-10}$, respectively.}
\label{figure:figure4}
\end{figure}


\end{document}